\documentclass{aa}  

\usepackage{graphicx}
\usepackage{txfonts}
\begin{document}

   \title{SDSS J1004+4112: the case for a galaxy cluster dominated by
    primordial black holes}

   \author{M.R.S. Hawkins
          \inst{1}}

   \institute{Institute for Astronomy (IfA), University of Edinburgh,
              Royal Observatory, Blackford Hill,
              Edinburgh EH9 3HJ, UK\\
              \email{mrsh@roe.ac.uk}}

   \date{Received September 15, 1996; accepted March 16, 1997}

 
  \abstract
   {}
   {This paper is aimed  at providing a plausible explanation for the
    large amplitude microlensing events observed in the cluster-lensed
    quasar system, SDSS J1004+4112.  The microlensed quasar images appear
    to lie well clear of the stellar population of the cluster, raising
    the possibility that the cluster dark matter is composed of compact
    bodies that are responsible for the observed microlensing.}
   {We established the exact structure of the
    difference light curves attributed to microlensing from photometric
    monitoring programmes in the literature.  We showed, based on measurements
    of surface brightness, that the probability of microlensing by stars in
    the cluster is negligible.   We relaxed our initial assumption that
    the cluster dark matter takes the form of smoothly distributed
    particles and we hypothesized that it is made up of compact bodies.
    We then used computer simulations of the resulting magnification
    pattern to estimate the probability of microlensing.}
   {Our results show that for a range of values for source size and lens
    mass, the large microlensing amplitude that is observed is consistent
    with the statistics obtained from the simulations.}
   {We conclude that providing the assumption of smoothly distributed dark
   matter is relaxed, the observed large amplitude microlensing can be
   accounted for by assuming that the cluster's dark matter is in the form of
   compact bodies of solar mass.  We further conclude that the most plausible
   identification of these bodies is that of primordial black holes.}

   \maketitle


\section{Introduction}

\begin{figure}
\begin{picture}(200,250)(-5,0)
\includegraphics[width=0.46\textwidth]{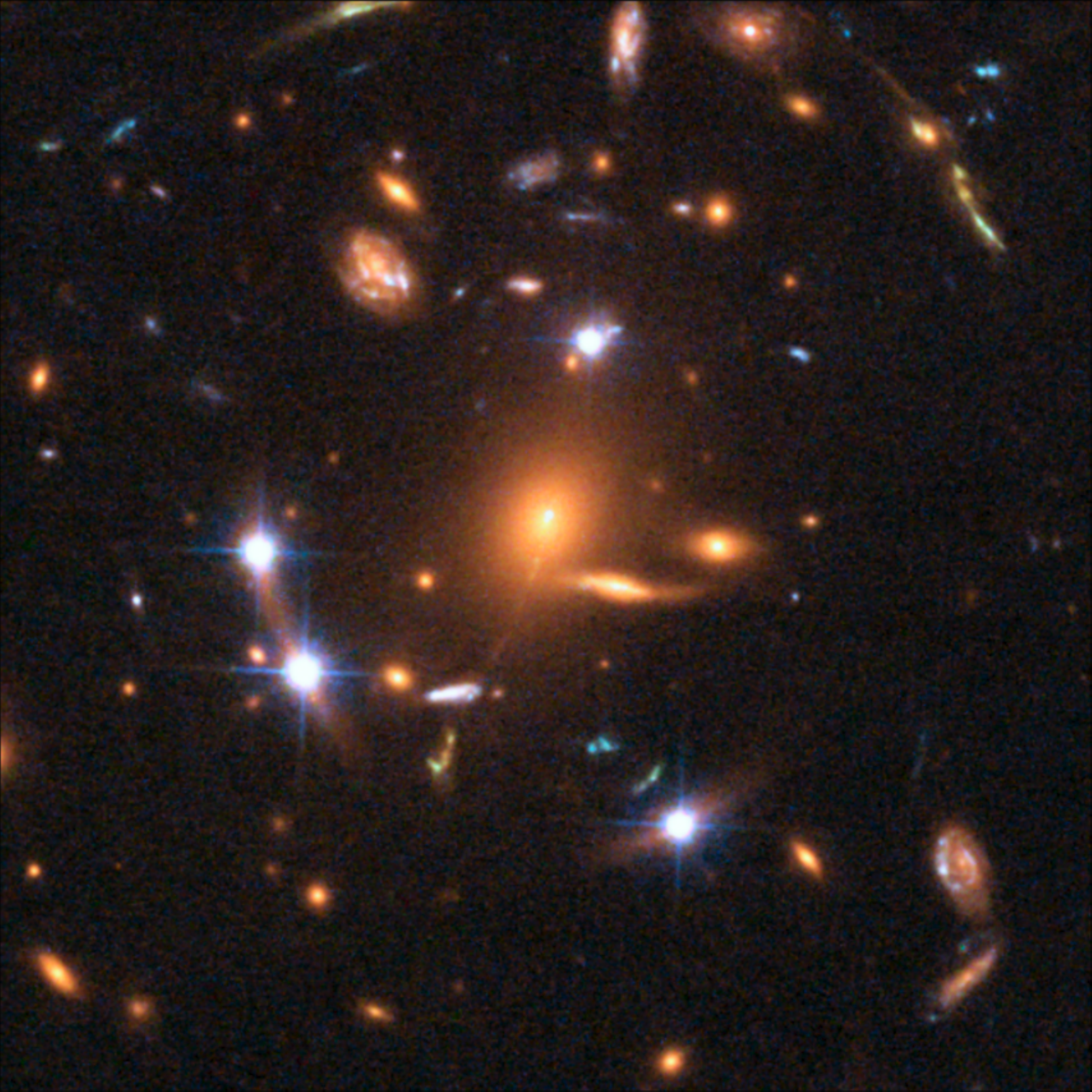}
\end{picture}
\caption{Hubble Space Telescope colour image of the lensed quasar
 SDSS J1004+4112, showing the lensing cluster.  The frame is
 approximately 30 arcsec on a side and north is up, east is to the
 left.}
\label{fig1}
\end{figure}

SDSS J1004+4112 is a rare example of a quasar that is lensed by a galaxy cluster
\citep{i03}.  The quasar at a redshift of $z_s = 1.734$ is split into five
images by a small galaxy cluster at $z_d = 0.68$.  The overall
configuration of the quasar images and galaxies in the lensing cluster is
shown in the {\it Hubble} Space Telescope (HST) composite image in
Fig.~\ref{fig1}.  In this paper, we follow the notation of
\cite{i03} for the identification of the quasar images A-D.   The system
has been extensively monitored photometrically \citep{f07,f08,f16}, with a
view to establishing time delays between the quasar images.  This has been
successful for delays between images A, B, and C, but the long interval for
image D has yet to be measured.  These monitoring programmes were
primarily motivated by the need to improve the constraints on the mass
modelling of the system, but in addition, an important result was the
detection of microlensing of the the A-B image pair \citep{f07}, which was
subsequently confirmed for all images \citep{f16}.  In addition to the
differential variation of the brightness of the images, microlensing has
also been detected by comparing brightness variations in different
photometric passbands \citep{r09} or between spectral lines and the
continuum \citep{m12}.  These measures also provide a zero-point for the
photometric difference light curves.  The other main focus of works on
SDSS J1004+4112 has been the measurement of the size of the quasar
accretion disc \citep{f08,f16}.  This is an important parameter for
estimating the expected microlensing amplitude and we discuss it in more
detail in Section~\ref{sec4}.

\begin{figure}[ht]
\centering
\begin{picture} (0,585) (150,5)
\includegraphics[width=0.9\textwidth]{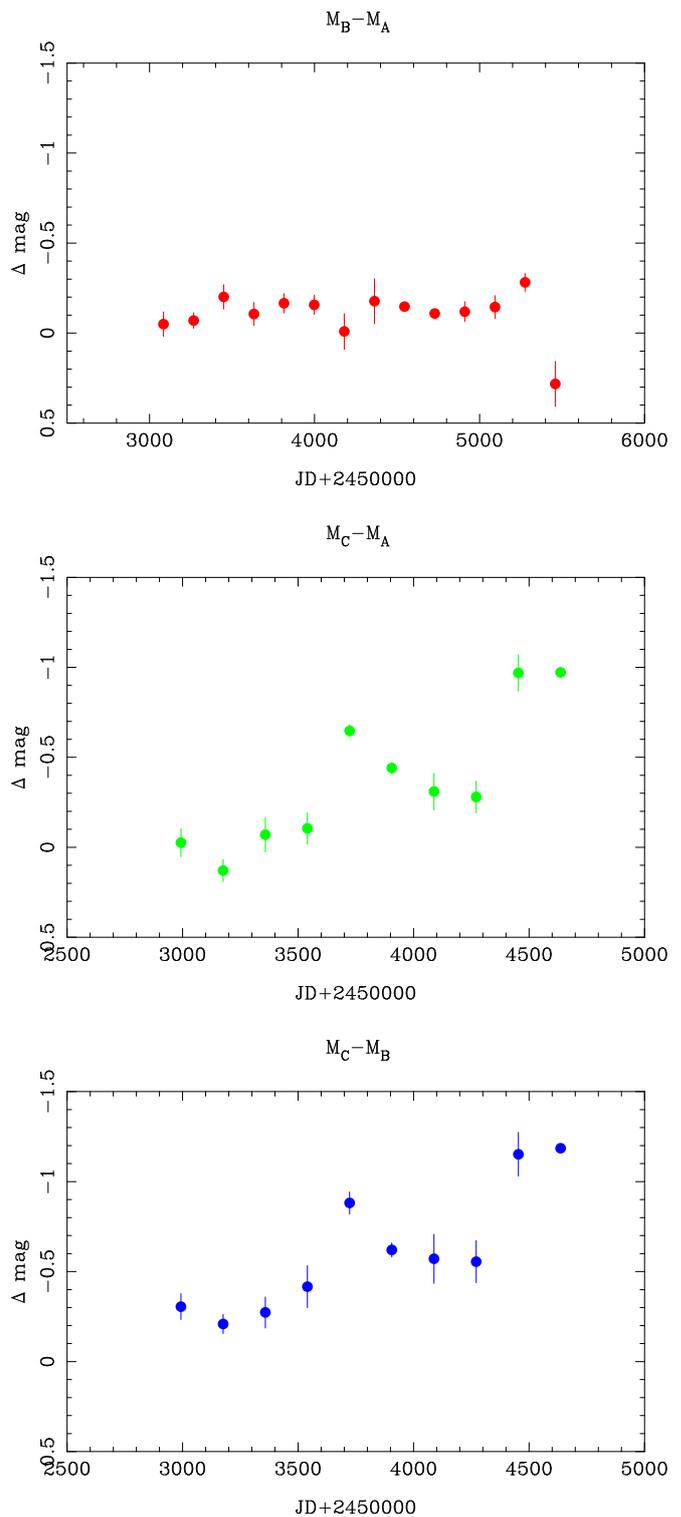}
\end{picture}
\caption{Difference light curves for images A, B, and C of the
 gravitationally lensed quasar SDSS J1004+4112.  The differential
 variations of  the images are attributed to microlensing.  The
 derivation of the zero-point is described in the text.}
\label{fig2}
\end{figure}

The main purpose behind this paper is to investigate the source of the
observed microlensing.  A cursory examination of the lensing cluster
illustrated in Fig.~\ref{fig1}, and in particular, the dominant central
galaxy, suggests that the stellar population has effectively become
negligible well before the radial distance of the quasar images close to
the Einstein ring is reached.  In this paper, we use measures from deep
HST frames in the infrared to estimate the probability of microlensing by
stars in the central cluster galaxy, based on the assumption that the dark
matter in the galaxy halo, as measured from modelling the lens system, is
smoothly distributed.  The chance superposition of dwarf galaxies on the
quasar images is another possibility that may account for the
microlensing.  To quantify this effect, we measured the optical depth to
microlensing over a large area of the cluster where there is a population
of dwarf galaxies, in both the infrared and visual wavebands.  From this
analysis, again assuming that the dark matter is smoothly distributed, we
conclude that it is very unlikely that the observed microlensing is caused
by stars.

If the assumption that the dark matter is smoothly distributed is
abandoned and, instead, we assume that it takes the form of compact bodies, the
probability that the quasar images are microlensed is greatly increased.
We quantify this unsurprising result with computer simulations of typical
magnification patterns for a population of solar mass bodies with
convergence and shear equal to those found from lensing models in the
vicinity of the quasar images.  The main conclusion of this paper is that
the probability that the observed microlensing in the cluster lens
SDSS J1004+4112 can be attributed to stars is very small.  If the
assumption of smoothly distributed dark matter is relaxed and we assume the cluster
halo is composed of stellar mass compact bodies, then the observed
microlensing is consistent with the statistics of the simulations.  We
further conclude that the most plausible candidates for the compact bodies
are primordial black holes of around one solar mass.  Throughout the paper,
we assume $\Omega_M = 0.27$, $\Omega_\Lambda = 0.73,$ and
$H_0 = 71$ km s$^{-1}$ Mpc$^{-1}$ for cosmological calculations.

\section{Light curves}
\label{sec2}

The main motivation behind this paper is to understand how quasar images
in an apparently sparse stellar environment can show strong variations
in brightness due to the effects of microlensing.  In this section, we
make use of extensive observations from the literature to establish the
amplitude of the microlensing variations.  After the discovery of
SDSS J1004+4112 \citep{i03}, the first light curves to be published
\citep{f07} showed variations in images A, B, C, and D over a three-year
period.  The time delay between image A and B was measured to be 38 days,
and when this delay was applied to the light curves, the flux ratio between
the two images was seen to vary with time, which was interpreted as a
detection of microlensing.  The light curves were extended for a further
year by \cite{f08}, who determined the time interval for image C which
they showed leads image A by $821.6 \pm 2.1$ days, and refined the time
delay between images A and B to $40.6 \pm 1.8$ days.  An additional four
years of photometric monitoring were added by \cite{f16}, who found a
strong microlensing signal for image C.

An alternative approach to measuring microlensing is based on the idea
that light in the infrared or from the broad or narrow line region of the
quasar comes from an area that is too large to be significantly microlensed.
In this case, photometry of the quasar images will yield the true flux
ratios resulting from the mass distribution and alignment of the lensing
galaxy or cluster.  By contrast, the structure of the accretion disc in
the UV rest frame are sufficiently compact that microlensing by compact
stellar mass bodies is a possibility.  By comparing the flux ratios of the
images in these two regimes, the effects of microlensing can be detected
in a single epoch.  This approach can also be used to determine the
zero-point for light curves obtained from photometric monitoring
programmes.  After allowing for time lags in light arrival time, the
magnitude difference between two quasar images removes the effects of
intrinsic variability.  The resulting light curve of magnitude difference
as a function of time provides a useful measure of microlensing, but it only
shows relative changes in magnitude difference.  The measures of intrinsic
flux differences described above can provide a zero-point for these
difference light curves to give the true magnitude difference due to
microlensing as a function of time.

The comparison of line strength with continuum flux was used to look
for microlensing in SDSS J1004+4112 by \cite{m09}.  Based on observations
made in 2004, they detected no microlensing in the A-B image pair, but they did detect a
strong microlensing amplification of 0.45 magnitudes for A-C.  They also
used these data to calculate the zero-points for the B-A and C-A difference
light curves, finding $(M_B-M_A)_0 = 0.50$ and $(M_C-M_A)_0 = 0.19$.  The
value for B-A was later refined by \cite{m12} to
$(M_B-M_A)_0 = 0.52 \pm 0.07$.  Based on their observations in 2008, \cite{m12}
also detected a microlensing signal for the A-B image pair of 0.2
magnitudes.  In addition, \cite{m09} compared their spectroscopic results
with flux ratios in the mid-infrared from \cite{r09} and concluded that
differences between extinction in the mid-infrared and optical where the
light curves are measured make spectroscopic measures of flux ratios the
preferred choice.  On this basis we follow \cite{m09} and \cite{m12} and
we use their spectroscopic zero-points in this paper.

To measure the amplitude and illustrate the morphology of microlensing
variations, we made use of the photometry of \cite{f07}, \cite{f08}, and
\cite{f16}.  Firstly, the epoch of each observation was corrected for time
lag due to the difference in time travel for light from the three images
using the results of \cite{f08}.  They find that image B leads image A by
$40.6 \pm 1.8$ days and image C leads image A by $821.6 \pm 2.1$ days.
Using these data, we converted all photometric observations to the
epoch of image A.  In order to emphasize the long-term trends in the data,
we binned them in half-yearly intervals and calculated the weighted
mean for each image.  In order to remove the effects of any intrinsic
variation of the quasar, we then subtracted the binned data for each epoch
to produce difference light curves for each pair of quasar images.
Figure~\ref{fig2} shows the difference light curves for the B-A, C-A, and
B-C image pairs.  Zero-points, as described above, from \cite{m09} and
\cite{m12} were applied to the light curves to give true measures of
the difference in microlensing between the two images.

The variations in the difference light curves in Fig.~\ref{fig2} show
that all image pairs are being microlensed, with amplitudes of 0.56, 1.10,
and 0.98 magnitudes for B-A, C-A, and C-B respectively.  This, in turn,
implies that at least two of the three images, and very probably all
three, are individually microlensed.

\section{The lensing cluster}
\label{sec3}

\begin{figure}
\centering
\begin{picture} (0,200) (150,0)
\includegraphics[width=0.6\textwidth]{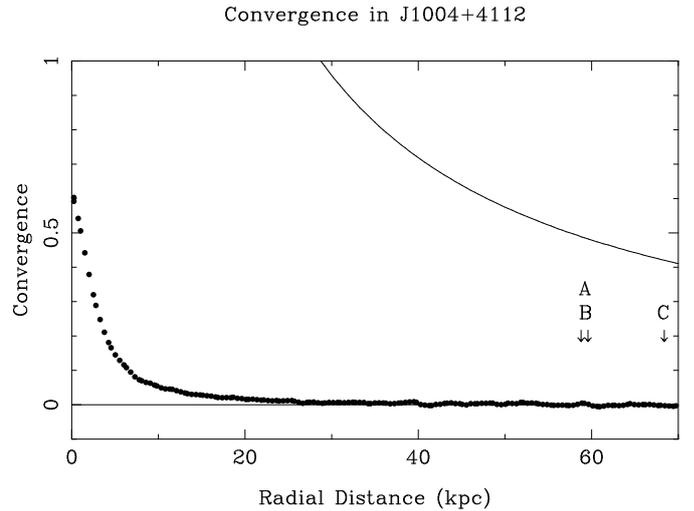}
\end{picture}
\caption{Convergence as a function of radial distance for the lensing
 cluster J1004+4112.  The solid line shows the total convergence,
 $\kappa$, from the lensing model and the filled circles show convergence,
 $\kappa_*$, for the stellar population of the lensing cluster.  The
 arrows indicate the positions of the quasar images.}
\label{fig3}
\end{figure}

\begin{figure*}
\centering
\begin{picture} (0,280) (255,0)
\includegraphics[width=0.49\textwidth]{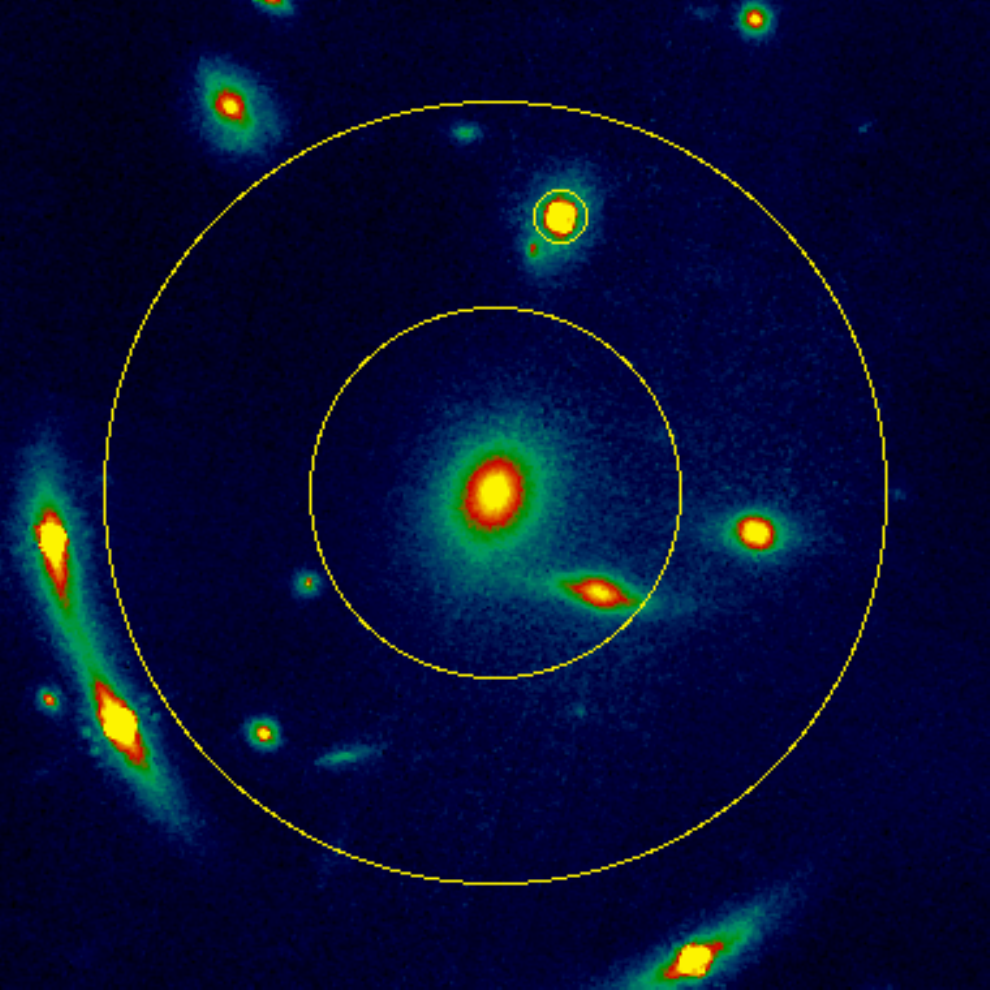}
\end{picture}
\begin{picture} (0,280) (-5,0)
\includegraphics[width=0.49\textwidth]{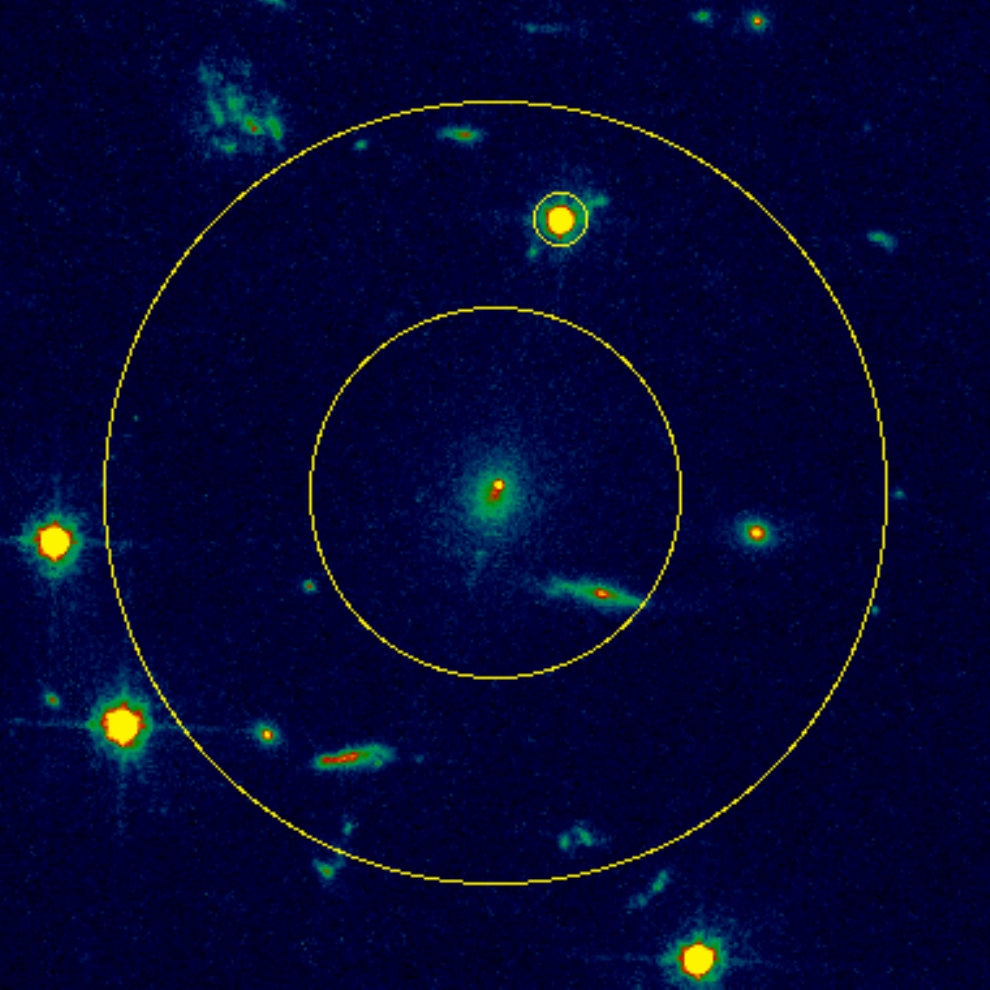}
\end{picture}
\caption{Deep HST frames of the area around the lensed quasar images of
 the cluster J1004+4112.  The photometric passbands are F160W (left-hand
 panel) and F555W (right-hand panel), close to the $H$ and $V$ bands,
 respectively.  The frames are approximately 15 arcsec on a side; 
 north is up, east to the left.  The frames illustrate the
 likelihood of the stellar populations of small cluster galaxies
 overlapping the quasar images and producing a microlensing signal.  The
 yellow circles define the boundary of the measurement of optical depth to
 microlensing for the dwarf cluster galaxies.  The three bright
 star-like images outside the outer circle are, from the north, images B,
 A, and C.}
\label{fig4}
\end{figure*}

In this section, we measure the optical depth to microlensing $\tau$ as a
function of radius for the stellar population of the central dominant
galaxy in the lensing cluster, based on the assumption that the dark matter is
smoothly distributed. Here,  $\tau$ is defined \citep{s92} as

\begin{equation}
\tau = \frac{\kappa_*}{1-\kappa_c}
\label{eqn1}
,\end{equation}

\noindent
where $\kappa_*$ and $\kappa_c$ are the surface mass densities in compact
bodies and smoothly distributed matter, respectively, in units of the
critical surface mass density $\Sigma_{cr}$, and

\begin{equation}
\Sigma_{cr} = \frac{c^2}{4 \pi G} \frac{D_q}{D_g D_{gq}}
\label{eqn2}
.\end{equation}

\noindent
Here, $D_g$, $D_q$, and $D_{gq}$ are the angular diameter distances to the
lensing galaxy, to the quasar, and from the galaxy to the quasar
respectively. To calculate $\tau$, we need values for $\kappa_*$ and
$\kappa_c$.

The procedure we follow is to start by measuring the surface brightness
as a function of radius from {\it Hubble} Space Telescope (HST) frames in
the F160W passband (close to the $H$-band).  The choice of this band is to
maximize the contrast between the blue light of the quasar images and the
largely red light of the galaxy stellar population.  The surface
brightness profile is then combined with the stellar mass-to-light ratio
to give the surface mass density profile.  When this is converted to units
of critical density, it traces $\kappa_*$, the convergence due to the stellar
population, as a function of distance from the galaxy centre.  To evaluate
$\tau$, we also need values for $\kappa_c$, the smoothly distributed dark
matter, which we obtain from mass models of the gravitational lens system.

The measurement of the surface brightness is more straightforward for
SDSS J1004+4112 than for most quasar systems, where the images are
embedded in the lensing galaxy and point spread function subtraction
techniques are necessary for measuring the uncontaminated galaxy profile
\citep{b97,h20}.  The wide separation of the images resulting from the
large mass of the lensing cluster puts them well clear of the central
dominant galaxy and enables a straightforward measurement of surface
brightness.  The first step was to measure a deep HST frame in the F160W
passband using standard photometric techniques to give a galaxy profile of
counts per square arcsecond as a function of distance from the galaxy
centre.  These measures were then converted to surface brightness using
the relevant HST photometric zero-point.  

The next step was to convert the surface brightness profile to units of
stellar luminosity per square arcsecond and apply $K$ and evolutionary
corrections from \cite{p97} to give a zero-redshift galaxy profile in
units of stellar luminosity in the $H$-band.  To convert stellar
luminosity to mass density, we use the value for mass-to-light ratio
derived by \cite {h20}.  As direct measures of stellar mass-to-light
ratios for galaxies are not normally feasible due to the pervasive
contribution of dark matter, this is based on dynamical measures for
globular clusters in the Galaxy and M31, where it is generally
assumed that the dark matter content is negligible.  The cluster masses
are derived from stellar velocity dispersions, and provide a direct
measure of stellar mass which can then be combined with photometry of the
clusters to give the stellar mass-to-light ratio.  An alternative approach
is to use population synthesis models to estimate the stellar
mass-to-light ratio in elliptical galaxies, which gives results consistent
with the dynamical measures.  There are a number of complicating issues
involved in these measurements that are discussed in detail by \cite{h20},
giving a final value of M/L$_H = 0.56 \pm 0.21$, which we adopt
in this paper.  The resulting surface mass density, when converted to
units of critical density using Eq.~\ref{eqn2}, gives $\kappa_*$ as a
function of distance from the central cluster galaxy.

To evaluate $\kappa_c$, the convergence from smoothly distributed
matter, we use values of total convergence $\kappa$ from lens
modelling of the quasar system.  Although there has been some difficulty
in accurately reproducing the positions of the quasar images and time
delays \citep{f08}, the values of convergence and shear from different
models are stable \citep{f08,f16}.  For this paper, we use values of
$\kappa$ and $\gamma$ from \cite{f16} for their singular isothermal sphere
plus external shear $({\rm SIS}+\gamma_e)$ model. $\kappa_c$ is now just
the difference between $\kappa$ and $\kappa_*$.

Figure~\ref{fig3} shows $\kappa_*$, the convergence resulting from the
stellar population of the dominant cluster galaxy, as a function of
distance from the galaxy centre.  Where the curve flattens out, a small
demagnification of $\kappa_* \sim -0.015$  is apparent, which we
removed with a median filter to give a clearer idea of the fluctuation
of the convergence around zero.  The positions of images A, B, and C are
marked at their distances from the centre of the galaxy.  Also
shown is the total convergence from the mass model of the quasar system
from \cite{f16}.  It can be seen that $\kappa_*$ becomes negligibly small
at around 25 kpc from the cluster centre and continues to fluctuate about
zero with an rms variation of $3.3 \times 10^{-4}$ out to 70 kpc, beyond
the furthest quasar image.

In order to convert $\kappa_*$ to optical depth to microlensing $\tau,$ we
make use of Eq.~\ref{eqn1}.  This equation is valid in the low optical
depth regime for $\tau \lesssim 0.1$ \citep{k97}, which is certainly
fulfilled for images A, B, and C.  From the lens model of \cite{f16}, the
total convergence $\kappa$ at the three image positions is 0.48, 0.47, and
0.38 respectively -- all close to a half as might be expected.  Given
the very small values for $\kappa_*$ this implies
$\kappa_c \approx \kappa$, and so, from Eq.~\ref{eqn1}, we have
$\tau \approx 2 \kappa_*$, which is still a very small value.

Given the large values of $\kappa_*$ at the quasar image positions needed
to produce the observed microlensing, it was suggested by \cite{f08}
that microlensing might be due to a satellite galaxy rather than
intracluster stars.  This interesting possibility can be investigated by
estimating the integrated optical depth to microlensing over the area of
sky occupied by the quasar images, including the contribution from stars
in dwarf galaxies.  Figure~\ref{fig4} shows deep images from the
{\it Hubble} Space Telescope in the F160W ({\it H}) and F555W ({\it V})
bands.  A small population of dwarf galaxies is visible, mostly red and
more easily visible in the F160W frame, although a small proportion of
faint blue galaxies can also be seen.  There is no obvious sign of any
small galaxy intercepting the line of sight to one of the quasar images,
but it is possible that the scattered light surrounding each image,
together with the lensed host galaxy of the quasar, is masking the presence
of a dwarf galaxy.  On each of the frames, we marked two large circles
centred on the central cluster galaxy.  The inner circle is at 25 kpc,
where light from the central galaxy is lost in the noise, as illustrated in
Fig.~\ref{fig3}.  The outer circle is at a radius of 53 kpc, and has been
chosen to exclude images A, B, and C.  We use the area between the two
circles as a basis for the calculation of the optical depth to
microlensing for the dwarf galaxy population of the cluster.  The area
includes parts of small galaxies which are large enough to be clearly 
visible had they overlapped with a quasar image, along with image D (which
we excluded, marked by a small circle).  The convergence, $\kappa_*$, over
this area was measured from the F160W frame (as for the profile of the
central cluster galaxy) and integrated to give the optical depth to
microlensing.  The average value of $\tau = 0.0095 \pm 0.0014$ should be
seen as an upper limit to the probability of microlensing by dwarf
galaxies as it includes contributions from larger galaxies and applies to
an area closer to the cluster centre than the positions of the quasar
images, where we would expect the density of cluster galaxies to be
higher.  Also, as we point out in Section~\ref{sec2}, at least two of the
quasar images are microlensed, which reduces the probability that dwarf
galaxies are the cause of this to less than $10^{-4}$.

\begin{figure*}
\centering
\begin{picture} (0,280) (255,0)
\includegraphics[width=0.49\textwidth]{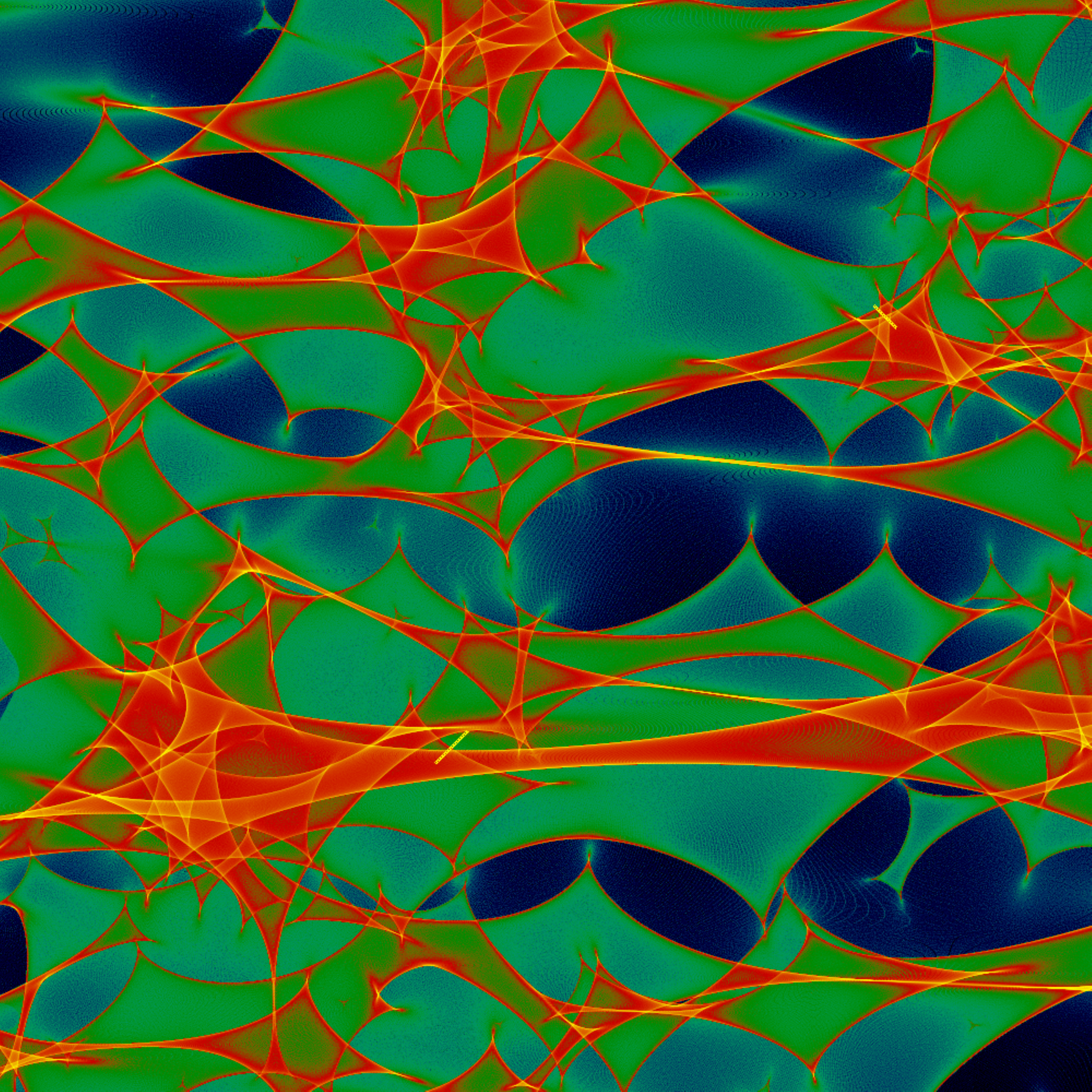}
\end{picture}
\begin{picture} (0,280) (-5,0)
\includegraphics[width=0.49\textwidth]{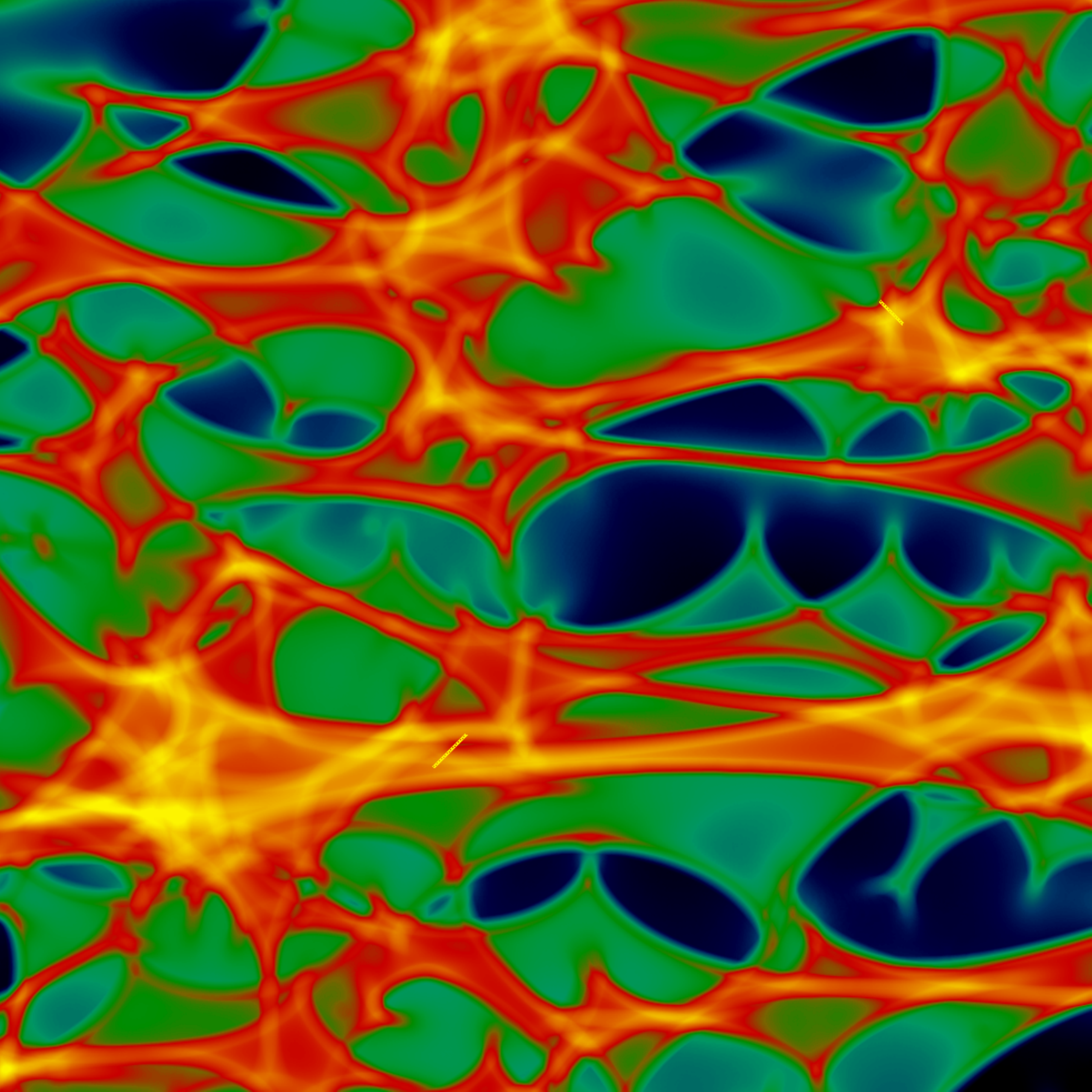}
\end{picture}
\caption{Microlensing magnification patterns for a population of 0.7
 M$_\odot$ bodies with parameter values of $\kappa_* = 0.39$,
 $\gamma = 0.47$ and a point source (left hand panel).  The right-hand
 panel shows the same simulation but for a source of radius one lt-day.
 Each frame has a side length of eight Einstein radii in the source plane.
 The yellow lines indicate tracks across the amplification pattern for the
 lengths of the B-A and C-A light curves in Fig.~\ref{fig2}.  A net
 transverse velocity of 600 km sec$^{-1}$ is assumed \citep{k86}.}
\label{fig5}
\end{figure*}

The very low probabilities for microlensing of the quasar images derived
thus far are based on the assumption that the dark matter in the cluster
is smoothly distributed, as would be expected if it were in the form of
elementary particles.  If this assumption is relaxed and, instead, we
assume dark matter is in the form of solar-mass compact bodies
\citep{c75,h93,j97,b18} the microlensing probabilities are greatly
increased.  In this situation, $\kappa_* \approx \kappa,$ and so, from the
lens modelling, we have $\kappa_* \sim 0.5$ in the vicinity of the quasar
images.  This puts the value of $\tau$ well outside the low optical depth
regime, as defined by \cite{k97}, and into a region where microlensing
probabilities can only be estimated from computer simulations of
microlensing amplification patterns.  A further motivation for estimating
microlensing probabilities from computer simulations is that they can
include the effects of shear and source size and take into account the
observed amplitude of microlensing.

\section{Microlensing simulations}
\label{sec4}

\begin{table*}[t]
\caption{Microlensing probabilities for SDSS J1004+4112}
\label{tab1}
\begin{center}
\vspace{5mm}
\begin{tabular}{ccccccccccrr}
\hline\hline
 & & & & & & & & & & & \\
 \multicolumn{2} {c} {Image C} &
 \multicolumn{2} {c} {Image B} & M$_{lens} $ &
 $R_D$ & $R_D/r_E$ & \multicolumn{2} {c} {Light curve} & Trials &
 $N(\Delta\!>\!\Delta m)$ & $P(\Delta m)$ \\
 $\kappa_*$ & $\gamma$ & $\kappa_*$ &
 $\gamma$ & $M_\odot$ & lt-day & & years & $\Delta m$ & & & \\
 & & & & & & & & & & & \\
\hline
 & & & & & & & & & & & \\
 0.38 & 0.00 & 0.47 & 0.00 &
               0.7 & 0 & 0.00 & 4.5 & 0.98 & $10^6$ & 230958 & 0.231 \\
 0.38 & 0.33 & 0.47 & 0.39 &
               0.7 & 0 & 0.00 & 4.5 & 0.98 & $10^6$ & 321324 & 0.321 \\
 0.38 & 0.33 & 0.47 & 0.39 &
               0.7 & 1 & 0.07 & 4.5 & 0.98 & $10^6$ & 100519 & 0.101 \\
 0.38 & 0.33 & 0.47 & 0.39 &
               0.3 & 1 & 0.11 & 4.5 & 0.98 & $10^6$ & 179594 & 0.180 \\
 & & & & & & & & & & & \\
\hline
 & & & & & & & & & & & \\
 \multicolumn{2} {c} {Image A} &
 \multicolumn{2} {c} {Image B} & & & & & & & \\
 & & & & & & & & & & & \\
\hline
 & & & & & & & & & & & \\
 0.48 & 0.57 & 0.47 & 0.39 &
               0.7 & 1 & 0.07 & 6.5 & 0.56 & $10^6$ & 598096 & 0.598 \\
 0.48 & 0.57 & 0.47 & 0.39 &
               0.3 & 1 & 0.11 & 6.5 & 0.56 & $10^6$ & 769067 & 0.769 \\
 & & & & & & & & & & & \\
\hline   
\end{tabular}
\end{center}
\end{table*}

In this section we present simulated amplification patterns for values of
$\kappa$ and $\gamma$ close to the quasar image positions of
SDSS J1004+4112 from the mass models of \cite{f16}.  The idea is to
explain the observed large amplitude microlensing and to this end, we
relax the commonly held assumption that dark matter takes the form of
smoothly distributed elementary particles and, instead, we allow for the
possibility that it is actually in the form of solar-mass compact bodies.  The
simulations were carried out using the ray tracing software of \cite{w99},
with values of $\kappa$ and $\gamma$ from \cite{f16}.  The left-hand panel
of Fig.~\ref{fig5} shows the simulation for values of $\kappa$ and
$\gamma$ at the position of image B, where a point source is assumed.

To quantify microlensing probabilities in this high optical depth regime,
we first superimpose a track of the length of the difference light curve,
as shown in Fig.~\ref{fig2}, on the amplification patterns for each of the
two images, in random position and orientation.  The magnification change
along each track can then be extracted to give simulated light curves in
magnitudes and subtracting one curve from the other gives a simulated
difference light curve.  By repeating this procedure a number of
times and counting the frequency with which the amplitude of the simulated
light curve exceeds that of the observed one, the likelihood that the
observed amplitude is caused by microlensing can be estimated.  To
illustrate this, in Fig.~\ref{fig4}, two such tracks are superimposed in
arbitrary positions and orientations of lengths corresponding to the
difference light curves for C-B and B-A.

The results from the microlensing simulations are shown in
Table~\ref{tab1}.  The first four columns show the values for $\kappa$ and
$\gamma$ used in the simulations for each image pair, taken from \cite{f16}.
The convergence is labelled $\kappa_*$ as we are assuming that the dark
matter is composed of compact bodies.  Other columns give the assumed lens
mass, quasar disc size in units of light days and Einstein radius, the
length of the difference light curve in years, and the observed amplitude,
$\Delta m$ in magnitudes, from the difference light curves in
Fig.~\ref{fig2}.  The final columns give the number of trials as described
above, the number of times, $N(\Delta>\Delta m)$, that the observed
amplitude is exceeded in the simulations, and the resulting probability of
microlensing.

The first two rows in Table~\ref{tab1} illustrate the effect of shear,
which has, so far, not been part of our optical depth calculations.  Broadly
speaking, shear smears the magnification pattern along a preferred axis,
making large amplifications less likely to be achieved, but small ones
more so.  The effect on the microlensing statistics depends on a number of
factors, including the amplitude of the observed light curve and the
source size, as well as the values of $\kappa_*$ and $\gamma$.  For the
data in Table~\ref{tab1}, shear increases the microlensing probability by
about 40\%.

So far in the simulations, we have assumed a point source but, in practice,
the quasar disc radius is a non-negligible fraction of the Einstein
radius of the lenses.  This means that the source is effectively resolved
by the lenses and the magnification reduced.  For very large sources, the
magnification in magnitudes is reduced asymptotically as
$\delta m \propto \theta^{-1}$, where $\theta$ is the source radius
\citep{r91}.  For smaller sources we must use microlensing simulations
and the third row of Table~\ref{tab1} shows the results for a quasar disc
radius of 1 lt-day.  A lens mass of $0.7 M_{\odot}$ is assumed, giving
a disc radius of 0.07 in units of lens Einstein radius.  It can be seen that this substantially reduces the probability of reproducing the observed
amplitude of the C-B difference curve.  We note that for a source size of
four lt-day, the probability becomes negligible.

In microlensing statistics, the mass of the lenses plays a significant
part.  Although the magnification patterns themselves do not depend on
mass, they are scaled to the size of the Einstein radius and, thus, the
area of sky they cover increases with mass.  This has two consequences:
1) the length of the light curves tracked across the
magnification patterns will be longer for smaller masses, making the
detection of a microlensing event more likely; 2) the size of the
accretion disc relative to the Einstein radius will be larger for smaller
masses.  As discussed above, this has the effect of reducing the
amplification and, hence, making microlensing events of a given magnitude
less likely.  These two effects work in opposite directions, but we have
found that, for the most part, the effect of a longer light curve relative
to the amplification pattern dominates and the frequency of microlensing
events increases with decreasing lens mass.  In the fourth row of
Table~\ref{tab1}, the lens mass is changed from $0.7 M_{\odot}$ to
$0.3 M_{\odot}$, keeping the other parameters unchanged, and we can see
that the microlensing probability increases somewhat.

So far, we have confined our attention to the C-B difference light curve,
which has a large amplitude of 0.98 mag and a relatively short run of
data due to the small overlap of the B and C light curves when adjusted
for time lag.  The B-A curve is longer with a smaller amplitude and in
the fifth and sixth rows of Table~\ref{tab1}, we show the microlensing
probabilities from the simulations.  To enable a direct comparison, the
same values for lens mass and source radius are used as in the third and
fourth rows for the C-B curve.  As might be expected, the likelihood of
microlensing is much greater for both assumed lens masses but, again, it is
relatively more probable for lower mass lenses.

\section{Discussion}
\label{sec5}

The objective of this paper is to provide an explanation for the
strong microlensing events observed in SDSS J1004+4112 for quasar images
which appear to lie completely clear of any significant distribution of
stars.  Firstly, we use
deep HST frames in optical and infrared bands to put precise limits on the
surface density of stars in the vicinity of the quasar images from the
dominant cluster galaxy and from chance superpositions of otherwise
undetectable dwarf galaxies.  For these measurements, we assume that the
cluster dark matter is in the form of smoothly distributed elementary
particles and find a negligible probability that the observed microlensing
can be attributed to stars.  In the second part of the paper, we relax the
assumption that dark matter is smoothly distributed and allow for the
possibility that it takes the form of compact bodies of around one solar
mass.  We then make use of computer simulations of the amplification
patterns around the microlensed quasar images, based on values for shear
and convergence from lens models of the system, to estimate the likelihood
that the observed microlensing amplitudes will be seen in simulated light
curves.  We find that the observed light curves are consistent with
microlensing by a dark matter halo made up of compact bodies of around a
solar mass.

It is interesting that the early photometric monitoring of SDSS J1004+4112
showed only a moderate microlensing amplitude of around 0.2 mag between
images A and B over a three-year period \citep{f08}, which is perhaps
sufficiently small to be accounted for by unidentified systematic effects.
However, the measurement of an 822 day time delay for image C \citep{f08},
along with four more years of photometric monitoring \citep{f16}, revealed
a change to this quiescent picture.  During this period, image B showed a
sharp decrease in brightness by 0.6 mag, while with the overlap between
image C and images A and B now at around five years it became clear that
image C was being strongly microlensed with an amplitude of around 1.0
mag.  Such a large amplitude is rare in any lensed quasar system, but in
wide separation systems it is only matched by RXJ 1131-1231 \citep{h20}.
It is also worth noting that the light curves in Fig.~\ref{fig2} bear a
strong qualitative resemblance to light curves from microlensing
simulations (for example \cite{k86}).

For most quasar systems where microlensing is observed, the quasar images
are produced by a massive galaxy along the line of sight acting as a lens
and the stars in this lensing galaxy are held to be responsible for the
microlensing.  In the case of SDSS J1004+4112, the situation is different
as the lens is a small cluster with a central dominant galaxy.  The
stellar population of this central galaxy is perhaps the most obvious
place to look for the source of microlensing, but as we see from
Fig.~\ref{fig3}, the surface density of stars, or convergence, $\kappa_*,$
has already dropped to a negligible level at a distance of 25 kpc from the
cluster centre.  As the microlensed images lie at 60 kpc and beyond, it
seems fair to conclude that microlensing from the stellar population of
the central galaxy can be ruled out beyond any reasonable doubt.  This
still leaves the possibility of the chance superposition of dwarf cluster
galaxies on the microlensed quasar images.  We first note that inspection
of the deep HST frames in Fig.~\ref{fig4} gives no indication that a dwarf
galaxy of any significant size overlaps the microlensed quasar images.
However, the scattered light from the quasar images and the lensed host
galaxy in the F160W frame make it hard to be sure that very small dwarf
galaxies are not coincident with the lines of sight to the microlensed
images.  Measurement of the integrated optical depth to microlensing gives
a conservative estimate of the probability that the quasar images are
microlensed by dwarf galaxies, or, indeed, any other stellar structure,
and the resulting low value implies that this is very unlikely to be the
solution.  This suggests that we seek a different route to explain the
observed microlensing signal.

If we relax the assumption that the dark matter in the cluster is composed
of smoothly distributed particles, the situation changes completely.  The
total convergence, $\kappa$, in the vicinity of the quasar images is around
a half, and in the case where the dark matter is in the form of compact
bodies, this implies a very large optical depth to microlensing.  This is
a regime where the probability of microlensing can only be estimated from
computer simulations.  In Table~\ref{tab1} we present a number
of results based on different assumptions for lens mass and source size.
To start with, we use values for convergence and shear for images C and B,
and the track length and amplitude are from the C-B difference light curve.
The first row in Table~\ref{tab1} shows a simulation for a point source
and no shear.  Neither of these are realistic assumptions, but they are
included to help visualize the effects these parameters have on
microlensing statistics.  For small values of $\kappa$ and $\gamma$, the
effect of $\gamma$ can usually be ignored \citep{s92}, but for larger
values it most often leads to an increase in microlensing probability, as
illustrated in Table~\ref{tab1}.  A more important issue is the effect
of the source size.  We assumed one lt-day for the quasar disc radius
from \cite{f08}, which is close to their value from thin disc theory, but
only marginally consistent with the measures of half-light radius from
\cite{f16}.  It can seen from Table~\ref{tab1} that increasing the
source size substantially decreases the probability of microlensing, as
expected.

The effect of lens mass on microlensing is discussed in
Section~\ref{sec4}, where we point out that in general, decreasing the
lens mass results in an increasing probability of microlensing.  So far,
we have assumed a lens mass of $0.7 M_{\odot}$ on the basis that compact
bodies forming a dark matter halo would most likely be primordial black
holes \citep{h20} and the best estimate of their mass would be
$0.7 M_{\odot}$ \citep{b18}.  However, typical values for primordial black
hole masses are still very uncertain and so, in the fourth row of
Table~\ref{tab1}, we show the effect of reducing the lens mass from
$0.7 M_{\odot}$ to $0.3 M_{\odot}$.  This roughly doubles the microlensing
probability due to the longer projected track of the light curve on the
amplification pattern.  For the last two rows in Table~\ref{tab1}, we
repeat the simulations in rows 3 and 4, but for images A and B.  The
change in the values for $\kappa_*$ and $\gamma$ do not have much of an
effect, but the A-B difference light curve is significantly longer and has
a smaller amplitude, which makes microlensing events much more likely.

The objective behind carrying out the microlensing simulations is to
establish whether relaxing the assumption that dark matter is smoothly
distributed can provide a plausible explanation for the observed large
amplitude microlensing in the quasar system.  The simulations described
in Table~\ref{tab1} cover a range of parameter values. Mostly, the
observed microlensing amplitude is well within the expectation from the
statistics of the simulations and in no case is it ruled out at the 95\%
confidence level.  From this, we conclude that although the probability of
microlensing from the stellar population of the cluster is negligible,
if the dark matter is in the form of compact bodies, then the observed
microlensing signal is easily explained.

The amplitude of the C-B difference curve is exceptionally large at around
one magnitude, which is interestingly close to the A-C curve for
RXJ 1131-1231.  In this case, the probability of microlensing from the
observed stellar population is also negligible \citep{h20}, but we note
that relaxing the assumption of smoothly distributed dark matter results
in microlensing probabilities similar to those in Table~\ref{tab1}.

Over the last decade or so, there have been a number of studies designed
to put constraints on the mass range of compact bodies which might make up
the dark matter.  The resulting mass limits have been collated by
\cite{c20} and appear to leave little room for dark matter in the form of
compact bodies.  However, the authors are at pains to point out that the
constraints have varying degrees of certainty and all come with caveats.
The constraints that are particularly relevant for primordial black holes
have been examined by \cite{b18}, with the overall conclusion that
essentially all of the limits rely on assumptions about astrophysical
processes or cosmological structures which are hard to verify.  More
specific to the work in this paper, the surface density of compact objects
at the positions of the quasar images in gravitationally lensed systems
can be estimated from the statistics of brightness changes due to
microlensing. The most recent in a number of studies employing this
technique \citep{m17} use single-epoch measurements of microlensing for a
large sample of lensed systems to conclude that compact bodies are most
likely to form only around 20\% of the total material of the galaxy.  This
result appears to be inconsistent with the results presented in this
paper, but it depends upon a number of assumptions, including the
clumpiness of the dark matter halo \citep{c18}, the mass function of the
lenses \citep{g17}, and the size of the quasar accretion disc.  Perhaps
more importantly, the samples used for statistical estimation are largely
composed of small separation systems where the quasar images are deeply
embedded in the stellar component of the lensing galaxy.  In this case,
there is no reason to believe that the stellar population cannot account
for the observed microlensing, but in samples confined to wide separation
systems the probability of microlensing by the stellar population becomes
small \citep{p12,h20}.

\section{Conclusions}
\label{sec6}

The main question we ask in this paper is how the large
microlensing amplification observed in the images of the quasar system
SDSS J1004+4112 can be accounted for, given that the stellar population of
the galaxy cluster appears to be negligible in the vicinity of the quasar
images.  Firstly, we used surface brightness measures from deep HST images
in the infrared to quantify the optical depth to microlensing for stars in
the dominant cluster galaxy at the positions of the microlensed images.
For this measurement, we assumed that the dark matter in the cluster is
in the form of smoothly distributed elementary particles.  We also
investigated the possibility that chance superpositions of dwarf cluster
galaxies along the line of sight to the quasar images could result in the
observed microlensing.  Our conclusion is that based on this assumption,
there is a negligible probability of producing the observed microlensing
signal.

In the second part of the paper, we relax the assumption that the cluster
dark matter is smoothly distributed.  Instead, we allow for the
possibility that it is in the form of compact bodies of around a solar
mass.  In this case, the large optical depth to microlensing at the
positions of the quasar images means that the likelihood of microlensing
can only be estimated from computer simulations of amplification patterns.
The values of convergence and shear for the simulations were taken from
mass modelling of the lensed system and microlensing probabilities were
calculated by projecting the tracks of the light curves onto the
amplification patterns and recording the frequency with which the observed
amplitude was exceeded.  A number of different combinations of input
parameters were used to illustrate the effect of source size and lens mass
on the probability of microlensing.  The results show that in all cases,
the observed microlensing amplitudes are consistent with the statistics
from the simulated amplification patterns.

In the event that compact bodies make up the dark matter, their
identification has been the subject of some speculation.  Suitable
candidates must be non-baryonic, sufficiently compact to microlens the
quasar images, and have masses in a range compatible with the observed
microlensing timescales.  Primordial black holes are, perhaps, the only
known candidates which satisfy these constraints \citep{h20}.  They thus
appear to be the most likely source of the lenses responsible for the
microlensing in SDSS J1004+4112, and hence to make up the dominant
component of the dark matter of the cluster.

\end{document}